\documentclass{PoS}
\newcommand{\Dslash}{{\not}\kern-0.05em D}
\newcommand{\dslash}{{\not}\kern+0.1em\partial}
\newcommand{\nablaslash}{{\not}\kern+0.15em\nabla}

\newcommand{\tr}{\mathop{\rm tr}\nolimits}

\newcommand{\SU}{\mathop{\rm SU}}

\newcommand{\U}{\mathop{\rm {}U}}
\newcommand{\rmd}{{\rm d}}

\title{Overlap fermion in external gravity}

\ShortTitle{Overlap fermion in external gravity}

\author{\speaker{Hiroto So}\thanks{After October 1st in  2006, 
the e-mail address is changed to 
 so@phys.sci.ehime-u.ac.jp  .}\\
        Department of Physics, Niigata University, Ikarashi 2-8050, 
        Niigata, 950-2181, Japan\\
        E-mail: \email{so@muse.sc.niigata-u.ac.jp}}

\author{Masashi Hayakawa\\
        Department of Physics, Nagoya University, Nagoya 464-8602, Japan\\
        E-mail: \email{hayakawa@eken.phys.nagoya-u.ac.jp}}

\author{Hiroshi Suzuki\\
        Theoretical Physics Laboratory, RIKEN, Wako 2-1, Saitama 351-0198, Japan\\
        E-mail: \email{hsuzuki@riken.jp}}

\abstract{On a lattice, we construct an overlap Dirac operator which 
describes the propagation of a  Dirac fermion in external gravity. The
local Lorentz symmetry is manifestly realized as a lattice gauge symmetry, 
while the general coordinate invariance is expected to be restored only in the 
continuum limit. The lattice index density in the presence of a gravitational 
field is calculated.}

\FullConference{XXIV International Symposium on Lattice Field Theory\\
		 July 23-28 2006\\
		 Tucson Arizona, US}

\newcommand{\be}{\begin{equation}}
\newcommand{\ee}{\end{equation}}

\begin{document}

\section{Motivations}
Lattice gauge theory is not just a regularized theory but 
a real calculable method for infinite degrees of freedom.   
There are two successes  in the  theory;
~exact gauge symmetry on lattice and  nonperturbative calculation of gauge dynamics. 
Nevertheless,  lattice fermion problems such as doubling 
and the chiral symmetry realization  were unsolved.  
Afterwards,  a new approach by an overlap fermion or Ginsparg-Wilson fermion brings to us two successes
\cite{Ginsparg:1982bj,Neuberger:1998fp,Narayanan:1998uu};~involving 
an exact chiral symmetry\cite{Luscher:1998pq} and ~anomaly calculation  
\cite{Hasenfratz:1998jp,Kikukawa:1998pd,Adams:1998eg,Suzuki:1998yz,Fujiwara:2002xh}. 
One may have a question:~~Can we involve gravity interaction and 
calculate the chiral anomaly on lattice? 

Usual approaches of the lattice gravity are Regge calculus  and 
 random lattice method\cite{Sorkin:1975ah}.  
We  mainly treat  only an external gravity 
and  use a fixed lattice(the spacing and the structure). 
So, our setting is the following; 
the lattice is spacing-fixed and cubic-connected, 
The coordinate is defined  as just an integer $x$, which is no real scale, 
gravity interaction is involved as a spin connection, 
 $U(x,\mu)\in {\rm spin(4)}$. 
Another  gravity  field is  denoted by a vierbein $e_\mu^a(x)\in{\rm GL}(4,R)$. 
Different from ordinary gauge symmetry cases, we must consider a nontrivial 
hermiticity (a metric) condition. 
As the result,  the spin connection is connected with the vierbein field. 
A serious problem of our method is  the continuum limit. 
Although the local Lorentz invariance is kept as exact lattice invariance, 
the general coordinate invariance  is expected after the continuum limit.  
In the formulation, can we calculate  the chiral anomaly? 
This is not a trivial problem.  
In this talk, the formulation of lattice overlap fermion with an external gravity 
and the calculation  of the chiral anomaly are sketched. 
For a detail description, see ref.~\cite{Hayakawa:2006vd}. 
In addition, an idea for a dynamical gravity is presented. 

\section{Lattice formulation with local Lorentz invariance}
We start with a lattice Dirac operator which is defined in terms of the
nearest-neighbor forward covariant difference:
\begin{equation}
   \nablaslash\psi(x)\equiv\sum_\mu
   \gamma^\mu(x)\{U(x,\mu)\psi(x+\hat\mu)-\psi(x)\}.
\label{threexone}
\end{equation}
This behaves covariantly, i.e.,
$\nablaslash\psi(x)\to g(x)\nablaslash 
\psi(x)$, under the local Lorentz
transformation
\begin{eqnarray}
    &&\psi(x)\to g(x)\psi(x),\qquad
    \overline\psi(x)\to\overline\psi(x)g 
(x)^{-1},
\nonumber\\
    &&U(x,\mu)\to g(x)U(x,\mu)g(x+\hat 
\mu)^{-1},\qquad
    \gamma^\mu(x)\to g(x)\gamma^\mu(x)g 
(x)^{-1},
\end{eqnarray}
where $\gamma^\mu(x)\equiv\sum_a e^\mu_a 
(x)\gamma^a$ and
$g(x)\in\mathop{\rm spin}(4)$. The {\it naive\/}
continuum limit of $\nablaslash$ coincides with the Dirac operator in the
continuum, denoted by $\Dslash$. To ensure that the lattice index theorem holds
in lattice gauge theory, in addition to the Ginsparg-Wilson relation, the
$\gamma_5$ hermiticity of a lattice Dirac operator is very important. We are
therefore naturally led to attempt to clarify the meaning of hermiticity in the
presence of a gravitational field.

We introduce the inner product of two functions on the lattice with spinor
indices as
\begin{equation}
   (f,g)\equiv\sum_xe(x)f(x)^{T*}g(x),
\label{threexthree}
\end{equation}
by using the determinant of the vierbein, $e(x)$. The inner
product~(\ref{threexthree}) is a natural lattice counterpart of the general
coordinate invariant inner product in a curved space. Next, we  note
\begin{equation}
   (f,\nablaslash g)
   =-(\nablaslash^*f,g)
   -\sum_x
   \left\{\sum_\mu\nabla_\mu^*\{e(x)\gamma^\mu(x)\}f(x)\right\}^{T*}g(x),
\label{threexfour}
\end{equation}
where a lattice Dirac operator is defined in terms
of the backward covariant difference,
\begin{eqnarray}
   \nablaslash^*\psi(x)
   &\equiv&\sum_\mu
   e(x)^{-1}e(x-\hat\mu)
   U(x-\hat\mu,\mu)^{-1}\gamma^\mu(x-\hat\mu)U(x-\hat\mu,\mu)
\nonumber\\
   &&\qquad\qquad\qquad\qquad\qquad\quad\times
   \left\{\psi(x)-U(x-\hat\mu,\mu)^{-1}\psi(x-\hat\mu)\right\},
\label{threexfive}
\end{eqnarray}
and the covariant divergence of $e(x)\gamma^\mu(x)$ is written as 
\begin{equation}
   \sum_\mu\nabla_\mu^*\{e(x)\gamma^\mu(x)\}
   \equiv
   \sum_\mu\{e(x)\gamma^\mu(x)
   -e(x-\hat\mu)U(x-\hat\mu,\mu)^{-1}\gamma^\mu(x-\hat\mu)
   U(x-\hat\mu,\mu)\}.
\label{threexsix}
\end{equation}
(Note that $\gamma^a$ and $\gamma^\mu(x)$ do not commute with the link
variables~$U(x,\mu)$, i.e., $[\gamma^\mu(x),U(y,\nu)]\neq0$.)
The $\nablaslash^*$ also behaves covariantly under the local Lorentz
transformation and coincides with the continuum Dirac operator~$\Dslash$ in
the naive continuum limit. The equality~(\ref{threexfour}) shows that,
if we postulate
\begin{equation}
   \sum_\mu\nabla_\mu^*\{e(x)\gamma^\mu(x)\}=0,\qquad\hbox{for all $x$}
\label{threexeight}
\end{equation}
on the external vierbein and the link variables as the lattice counterpart of
the metric condition, then the~$\nablaslash^*$ in eq.~(\ref{threexfive}) is
precisely the minus hermitian conjugate of $\nablaslash$ in
eq.~(\ref{threexone}):
\begin{equation}
   \nablaslash^* =-\nablaslash^\dagger,\qquad
   (\nablaslash^*)^\dagger=-\nablaslash.
\label{threexnine}
\end{equation}

Having clarified the meaning of the hermitian conjugation, it is
straightforward to construct a lattice Dirac operator of overlap type
with the desired properties. We first define the Wilson-Dirac operator by
\begin{equation}
   D_{\rm w}={1\over2}
   \left\{\nablaslash+\nablaslash^*
   -{1\over2}(\nablaslash^*\nablaslash+\nablaslash\nablaslash^*)\right\},
\label{threexten}
\end{equation}
where the second term is the Wilson term in a curved space, which is hermitian
with respect to the inner product~(\ref{threexthree}). Then in analogy to the
case of lattice gauge theory, we define the overlap-Dirac operator from the
Wilson-Dirac operator as
\begin{equation}
   D=1-A(A^\dagger A)^{-1/2},\qquad A\equiv1-D_{\rm w}.
\label{threextwelve}
\end{equation}
The naive continuum limit of the operator~$D$ is $\Dslash$. It is easy to
confirm that the free Dirac operator does not suffer from species doubling.

From its construction~(\ref{threextwelve}) and the $\gamma_5$ hermiticity
of the Wilson-Dirac operator, we find that the operator~$D$
satisfies the Ginsparg-Wilson relation in the conventional form,
\begin{equation}
   \gamma_5D+D\gamma_5=D\gamma_5D, 
\label{threexfifteen}
\end{equation}
and that it is also $\gamma_5$ hermitian, i.e.,
\begin{equation}
   D^\dagger=\gamma_5D\gamma_5,
\label{threexeighteen}
\end{equation}
with respect to the inner product~(\ref{threexthree}).

\section{Calculation of lattice index density}
We have defined a lattice Dirac operator of overlap type that
describes the propagation of a single Dirac fermion in external
gravitational fields. Since the forms of the Ginsparg-Wilson relation and the
$\gamma_5$ hermiticity are identical to those in lattice gauge theory, we
can repeat the same argument for the index theorem in the latter theory.

A natural lattice action for the massless Dirac fermion in a curved space is
\begin{equation}
   S_{\rm F}=\sum_xe(x)\overline\psi(x)D\psi(x).
\label{threexnineteen}
\end{equation}
This action is invariant under the modified chiral transformation
\begin{equation}
   \psi(x)\to\left(1+i\theta\hat\gamma_5\right)\psi(x),\qquad
   \overline\psi(x)\to\overline\psi(x)\left(1+i\theta\gamma_5\right),
\label{threextwenty}
\end{equation}
where $\theta$ is an infinitesimal constant parameter and
$\hat\gamma_5\equiv\gamma_5(1-D)$, due to the
Ginsparg-Wilson relation~(\ref{threexfifteen}). The functional
integration measure is, however, not invariant under this transformation and
gives rise to a non-trivial Jacobian~$J$, where
\begin{equation}
   \ln J=-2i\theta\sum_x\tr\Gamma_5(x,x)
\label{threextwentytwo}
\end{equation}
and
\begin{equation}
   \Gamma_5(x,y)\equiv\gamma_5\left(\delta_{xy}-{1\over2}D(x,y)\right).
\label{threextwentythree}
\end{equation}
The operator~$\Gamma_5$ anti-commutes with the hermitian
operator~$H\equiv\gamma_5D$, again due to the Ginsparg-Wilson relation.
It is then easy to prove the lattice index theorem. The quantity
\begin{equation}
   \sum_x\tr\Gamma_5(x,x)=n_+-n_-,
\label{threexthirty}
\end{equation}
where $n_\pm$ denote the numbers of zero eigenmodes of the~$H$ with positive
and negative chiralities, respectively, is an {\it integer\/} even for finite
lattice spacings. This is the lattice index theorem in the presence of a
gravitational field.

We next consider the classical continuum limit of the density of the lattice
index~(\ref{threexthirty}), which is precisely the chiral anomaly. Although the
direct calculation is quite involved, we can resort to the following argument
which utilizes the topological nature of the lattice index. First we note that
in the expressions of the index, $\Gamma_5$ can be replaced by
$\Gamma_5e^{-{H^2/M^2}}$ with an arbitrary mass~$M$, because only zero
eigenmodes of $H$ contribute to the index. Namely, we have
\begin{equation}
   \sum_x\tr\Gamma_5(x,x)=\sum_x\tr\sum_y\Gamma_5e^{-H^2/M^2}(x,y)
   \delta_{yx}
   \equiv a^4\sum_x{\mathcal A}_5(x),
\label{threexthirtyone}
\end{equation}
where the lattice spacing~$a$ has been restored, and the index density has
been defined by
\begin{equation}
 {\mathcal A}_5(x)\equiv
   {1\over a^4}\tr\sum_y\Gamma_5e^{-H^2/M^2}(x,y)\delta_{yx}.
\label{threexthirtytwo}
\end{equation}
From this point, we consider the index density on a lattice with infinite
extent, as usual for the classical continuum limit. Then we have
\begin{eqnarray}
   {\mathcal A}_5(x) =\int_{\mathcal B}{\rmd^4k\over(2\pi)^4}\,
   e^{-ikx}\tr\left(\gamma_5-{a\over2}H\right)e^{-{H^2/M^2}}e^{ikx},
\label{threexthirtythree}
\end{eqnarray}
where ${\mathcal B}$ denotes the Brillouin zone.
Then using the fact that the lattice free Dirac operator does not possess
doubler's zero, and $e^{-H^2/M^2}$ acts as a suppression factor at the
boundary of the Brillouin zone for $a\to0$, we can show that
\begin{eqnarray}
   \lim_{a\to0}\mathcal{A}_5(x)
   &=&\int_{{\mathbb R}^4}
   {\rmd^4k\over(2\pi)^4}\,\lim_{a\to0}
   e^{-ikx}\tr\left(\gamma_5-{a\over2}H\right)e^{-{H^2/M^2}}e^{ikx}
\nonumber\\
  &=&\int_{{\mathbb R}^4}
   {\rmd^4k\over(2\pi)^4}\,
   e^{-ikx}\tr\gamma_5e^{\Dslash^2/M^2}e^{ikx},
\label{threexfortyfive}
\end{eqnarray}
because $\lim_{a\to0}H=\gamma_5\Dslash$ in the naive continuum limit. It is
interesting that, in this calculational scheme~\cite{Fujikawa:1998if}, the term
proportional to $aH/2$ does not contribute. In this way, we obtain the
index density {\it in the continuum theory}. Namely,
\begin{equation}
   \lim_{a\to0}a^4\sum_x\tr\Gamma_5(x,x)
   =\int_{M_4}\lim_{a\to0}{\mathcal A}_5(x)
   =\int_{M_4}
   \det\left\{{i\hat R/4\pi\over\sinh(i\hat R/4\pi)}\right\}^{1/2},
\label{threexfortysix}
\end{equation}
where the curvature 2-form is defined by
$(\hat R)_a{}^b={1\over2}R_{\mu\nu a}{}^b\,\rmd x^\mu\wedge\rmd x^\nu$.
Thus as expected from the absence of the species doubling
(which implies the correct number of degrees of freedom) and the topological
properties of the lattice index, we find that our formulation reproduces the
correct expression of the chiral $\U(1)$ anomaly in a gravitational field. This
demonstration can be regarded as a test of the restoration of the general
coordinate invariance in the classical continuum limit.

We can also generalize our construction to other Lorentz (reducible)
representations. By a similar argument to the above, we have, for the
spinor-vector representation,
\begin{equation}
   \lim_{a\to0}\sum_x\tr\Gamma_5(x,x)
   =\int_{M_4}
   \det\left\{{i\hat R/4\pi\over\sinh(i\hat R/4\pi)}\right\}^{1/2}
   \tr\{e^{-i\hat R/(2\pi)}\}
\label{fourxfive}
\end{equation}
and, for the bi-spinor representation
\begin{equation}
   \lim_{a\to0}\sum_x\tr\Gamma_5(x,x)
   =\int_{M_4}
   \det\left\{{2i\hat R/4\pi\over\tanh(i\hat R/4\pi)}\right\}^{1/2}.
\end{equation}
These are well-known expressions of the index for those representations.

\section{Summary}
We formulated  an overlap fermion on lattice with an external gravity  
and calculated  a chiral index density.  In the formulation, several essential points 
are listed as 
~(i) the local Lorentz symmetry  is realized as the exact lattice gauge symmetry; 
~(ii) the general coordinate invariance is restored after the continuum limit; 
~(iii) a hermiticity condition of conjugate difference operators 
is important;
~(iv) a Wilson-Dirac operator is constructed in advance, and  
~(v) an overlap-Dirac operator is constructed  and the  Ginsparg-Wilson 
relation  is verified;  
~(vi) the calculation of the chiral index density 
can be executed not only for Dirac fermions but also for other representations. 

Validity of this formulation depend on the continuum limit strongly. 
If this continuum limit  or the fixed point has the property of a general covariance, 
a next step is to involve the  dynamics of gravity.   
The limit is expected to hold a relation, 
\begin{equation}
   \langle e_\mu^a(x)e_\nu^a(x)\rangle=g_{\mu\nu}(x) ,
\end{equation}
\noindent
where $\langle \cdots \rangle$ means  an average under the dynamics of vierbein fields 
to fix a physical scale(distance).  
 Vierbein fields  generally express a connection between a curved space and the tangent space. 
Although we treat  only a lattice tangent space, 
a real curved space may be realized after the dynamics of vierbein fields and 
some smoothing of the tangent space.

\acknowledgments{
This work is supported in part by the 
Grants-in-Aid for Scientific Research Nos.  13135203, 
13135223, 17540242, 18034011 and~18540305 from 
the Japan Society for the Promotion of Science.}

\end{document}